# Toggling Near-field Directionality via Polarization Control of Surface Waves

Yuhan Zhong[1,2,#], Xiao Lin[1,2,#,*], Jing Jiang[3], Yi Yang[4], Gui-Geng Liu[5], Haoran Xue[5], Tony Low[6,*], Hongsheng Chen[1,2,*], Baile Zhang[5,7,*]

[1]*Interdisciplinary Center for Quantum Information, State Key Laboratory of Modern Optical Instrumentation, ZJU-Hangzhou Global Science and Technology Innovation Center, College of Information Science and Electronic Engineering, Zhejiang University, Hangzhou 310027, China.*
[2]*International Joint Innovation Center, ZJU-UIUC Institute, Zhejiang University, Haining 314400, China.*
[3]*School of Applied Science, Beijing Information Science and Technology University, Beijing 100192, China.*
[4]*Research Laboratory of Electronics and Department of Physics, Massachusetts Institute of Technology, Cambridge, Massachusetts 02139, USA.*
[5]*Division of Physics and Applied Physics, School of Physical and Mathematical Sciences, Nanyang Technological University, Singapore 637371, Singapore.*
[6]*Department of Electrical and Computer Engineering, University of Minnesota, Minneapolis, Minnesota 55455, USA.*
[7]*Centre for Disruptive Photonic Technologies, Nanyang Technological University, Singapore 637371, Singapore.*
[#]*These authors contributed equally to this work.*
[*]*Corresponding authors. Email: xiaolinzju@zju.edu.cn (X. Lin); tlow@umn.edu (T. Low); hansomchen@zju.edu.cn (H. Chen); blzhang@ntu.edu.sg (B. Zhang)*

Directional excitation of guidance modes is central to many applications ranging from light harvesting, optical information processing to quantum optical technology. Of paramount interest, especially, the active control of near-field directionality provides a new paradigm for the real-time on-chip manipulation of light. Here we find that for a given dipolar source, its near-field directionality can be toggled efficiently via tailoring the polarization of surface waves that are excited, for example, via tuning the chemical potential of graphene in a graphene-metasurface waveguide. This finding enables a feasible scheme for the active near-field directionality. Counterintuitively, we reveal that this scheme can transform a circular electric/magnetic dipole into a Huygens dipole in the near-field coupling. Moreover, for Janus dipoles, this scheme enables us to actively flip their near-field coupling and non-coupling faces.

## 1. Introduction

The directional excitation of light in nanophotonic waveguides is often denoted as the *near-field directionality*, which is of fundamental and technological importance to modern photonic science [1-9]. It enables many applications, such as nanorouters [10,11], nanopolarimeters [12], nanoscopic position sensing [13,14], near-field microscopy [1], and the development of on-chip information processing and complex quantum networks [3,15-19]. In general, it is achieved through the judicious design of either asymmetric waveguide structures [20-22] or complex excitation sources [7,23-28], such as the extensively-studied circularly polarized dipoles. Circular electric (magnetic) dipoles [5,25, 29-31] are featured with a spinning electric (magnetic) dipole moment. Correspondingly, their near-field directionality is determined by the spin-momentum locking and can be understood from the perspective of quantum spin Hall effect of light [32-35]. Notably, recent studies show that the near-field directionality can go beyond the rudimentary spin-momentum locking, if the excitation source is the composite electric and magnetic dipoles [23-25]. For example, Huygens and Janus dipoles are constructed by the orthogonal electric and magnetic dipoles, which are in phase and 90° out of phase to each other, respectively [25]. As a result, the near-field directionality of Huygens and Janus dipoles is related to the real and imaginary parts of the Poynting vector, respectively [26, 36-38]. Since the circular, Huygens and Janus dipoles have different constructions, they have distinct features in the near-field coupling, as shown in the pioneering work of Ref. [25]. Particularly, Ref. [25] shows that the Janus dipole has two distinct faces, namely the near-field coupling and non-coupling faces. Circular, Huygens and Janus dipoles are then considered as three elemental dipolar sources, since their construction only requires two dipole components, which are the minimum requirements for the directional constructive/destructive near-field interference [23,25]. However, once these dipolar sources are given, the near-field directionality is generally predefined. Studies on the active control of this exotic phenomenon have so far been exclusively focusing on the modulation of excitation sources, such as their helicity and incidence angle [5, 39-46]. Without resorting to the modulation of excitation sources, the active control of this phenomenon remains challenging.

Here we show that the near-field coupling is a combined result of the dipolar source and the polarization of surface waves, instead of solely determined by the dipolar source itself. The underlying reason is that the surface waves play the key role to extract the evanescent electromagnetic energy from the dipolar source, and their polarization unavoidably has a significant influence on the near-field coupling efficiency. To be specific, the near-field coupling efficiency depends on both the dipolar source and the intrinsic property of surface waves via $|\bar{p} \cdot \bar{E}^* + \bar{m} \cdot \bar{B}^*|^2$, where $\bar{p}$ ($\bar{m}$) is the electric (magnetic) dipole moment, $\bar{E}$ and $\bar{B}$ are the electric field and magnetic flux at the position of dipolar source, respectively. Correspondingly, we reveal a feasible scheme for the active control of near-field directionality through tailoring the polarization of surface waves, without resorting to the modulation of excitation sources. The polarization of surface waves is flexibly tunable, for example, through changing the chemical potential of graphene (e.g., electrostatic gating) in a graphene-metasurface waveguide. In addition, the polarization of surface waves in this waveguide is readily tunable in a broad frequency range, spanning the whole terahertz (THz) range. Remarkably, we find that by following this scheme, a Huygens dipole can be transformed into a circular electric/magnetic dipole in the near-field coupling. Such a counterintuitive and hidden correspondence between circularly polarized dipoles and Huygens dipoles sheds new light on the manipulation of near-field coupling. Moreover, we further find that for Janus dipoles, our scheme can be exploited to actively flip their near-field coupling and non-coupling faces. The real-time manipulation of light flow at the subwavelength scale revealed in this work might usher new active on-chip optical functionalities.

2. **Results and Discussion**

We begin by exploring the overlooked role of the polarization of surface waves in the near-field coupling

[Fig. 1(a, b)]. Without loss of generality, we consider an excitation source above a planar waveguide, which is parallel to the $xy$ plane and supports either TM (transverse magnetic, or *p*-polarized) or TE (transverse electric, or *s*-polarized) surface waves. To facilitate the conceptual demonstration, the excitation source [Fig. 1(a)] is composed of one magnetic dipole moment (i.e., $m_z = cp_0$) and two electric dipole moments ($p_z = ip_0$ and $p_x = p_0$), where $c$ is the speed of light in free space, and we set $p_0 = 1$ Coulomb-meter. If the waveguide supports TM surface waves, this source is equivalent to a circular electric dipole with $p_z = ip_0$ and $p_x = p_0$ in the near-field coupling [Fig. 1(a)], since the component $m_z$ does not excite any TM surface waves. In contrast, if the waveguide supports TE surface waves, this source becomes equivalent to a Huygens dipole with $m_z = cp_0$ and $p_x = p_0$ in the near-field coupling [Fig. 1(a)], because the component $p_z$ cannot excite TE surface waves. According to the definition of Huygens dipole [25], its magnetic and electric dipole components are orthogonal to each other and are required to satisfy the Kerker's condition of $m/p = c$ [2,25,47,48]. In short, the designed source [Fig. 1(a)] has distinct characters in the near-field coupling for the two different polarizations of surface waves, and it is thus denoted as the bi-state dipole below. Figure 1(a) depicts the conceptual idea of utilizing the dual role of the dipolar source and the polarization of surface waves to enable unprecedented tunability in the near-field coupling.

Similarly, we can also construct another bi-state dipole [Fig. 1(b)], which has $p_z = p_0$, $m_z = icp_0$ and $m_x = cp_0$, in the near-field coupling for different polarizations of surface waves. To be specific, this bi-state dipole in the near-field coupling is equivalent to a Huygens dipole with $m_x = cp_0$ and $p_z = p_0$ if the waveguide supports TM surface waves, while it becomes to a circular magnetic dipole with $m_z = icp_0$ and $m_x = cp_0$ if the waveguide supports TE surface waves [Fig. 1(b)]. Figure 1(a, b) jointly shows that tailoring the polarization of surface waves provides a viable way to actively switch the performances of these two bi-state dipoles in the near-field coupling. The physical depiction in Fig. 1 sets the stage for our work. In what follows, we seek a versatile waveguide platform which allows for the active control of the polarization of surface waves.

Figure 2 illustrates a hybrid graphene-metasurface waveguide, designed to accommodate both TE and TM surface waves within the THz range (< 3 THz), where the switching of polarization is achieved through tuning the chemical potential of graphene. The graphene-metasurface waveguide is constructed by one layer of metal-based metasurface and graphene multilayers; herein the layer number of graphene is chosen to be $N = 7$ [Fig. 2(a)]. The stacking order of these layers is quite flexible in this waveguide [Fig. S1], and these layers are separated by a dielectric spacing, which has a relative permittivity of 3.9 (e.g., SiO$_2$) [49,50] and a deep-subwavelength thickness (e.g., 20 nm) [Fig. 2(a)]. The ultrathin metal-based metasurface is constructed by two-dimensional periodic copper patches, which are separated by air gaps with a width of $g = 5$ nm, and it has a pitch of $D = 5.2$ μm along both $x$ and $y$ directions [Fig. 2(a)]. Due to the deep-subwavelength size of unit cells, the metal-based metasurface can be effectively modelled with a surface conductivity of $\sigma_{\text{meta}}$ [51,52]. Moreover, this graphene-metasurface waveguide can be readily modelled with an effective surface conductivity of $\sigma_{\text{total}} = N\sigma_{\text{gra}} + \sigma_{\text{meta}}$ [Fig. S1], since the designed waveguide has a deep-subwavelength thickness, where $\sigma_{\text{gra}}$ is the surface conductivity of each graphene layer. The detailed verification of this approach is shown in Supplementary Note 1.

In principle, the polarization of surface waves in the graphene-metasurface waveguide is determined by the sign of $\text{Im}(\sigma_{\text{total}})$ [53-59]. That is, this waveguide in free space only supports TM surface waves if $\text{Im}(\sigma_{\text{total}}) > 0$ or TE surface waves if $\text{Im}(\sigma_{\text{total}}) < 0$ [53-59]. For our proposed geometry, we have $\text{Im}(\sigma_{\text{meta}}) < 0$ for the meta-based metasurface [Supplementary Note 1] and $\text{Im}(\sigma_{\text{gra}}) > 0$ for the doped graphene. Since $\text{Im}(\sigma_{\text{gra}})$ is sensitive to $\mu_c$, the sign of $\text{Im}(\sigma_{\text{total}})$ can be actively toggled by tuning the chemical potential $\mu_c$ of graphene. As an example, $\text{Im}(\sigma_{\text{total}}) = 59G_0$ if $\mu_c = 0.3$ eV but $\text{Im}(\sigma_{\text{total}}) =$ $-118G_0$ if $\mu_c = 0$ eV at 2.5 THz [Fig. S2], where $G_0 = e^2/4\hbar$ is the universal optical conductivity. Hence, this proposed graphene-metasurface waveguide can be one realization, and any waveguide design that allows for active toggling of the sign of $\text{Im}(\sigma_{\text{total}})$ would qualify.

In conjunction to the above, we discuss the second defining feature of our design. As shown by the dispersion curves in Fig. 2(b), the graphene-metasurface waveguide supports only TM surface waves if $\mu_c = 0.3$ eV or TE surface waves if $\mu_c = 0$ eV within the THz range. Especially, the dispersion curves for TE and TM surface waves have a crossing at 2.5 THz [Fig. 2(b)]. To be specific, we have $k_\parallel^{\text{TE}} = k_\parallel^{\text{TM}} = 1.7k_0$ at $\omega_0/2\pi = 2.5$ THz, where $k_\parallel^{\text{TE}}$ ($k_\parallel^{\text{TM}}$) is the in-plane wavevector for TE (TM) surface waves, $k_0 = 2\pi/\lambda_0$ and $\lambda_0 = 2\pi c/\omega_0 = 120$ μm. Therefore, the *accidental degeneracy* at this frequency enables the selective excitation of surface waves with different polarizations, while simultaneously maintains their same in-plane wavelengths (namely their same spatial confinements). Remarkably, the frequency of this accidental degeneracy is readily tunable within the terahertz regime as shown in Fig. S3.

We show in Fig 2(c, d) the field distribution of excited surface waves by putting an electric dipole with $\bar{p} = \hat{x}p_x$ above the graphene-metasurface waveguide at 2.5 THz. Remarkably, the spatial field distributions of excited surface waves simply rotate by 90 degrees when we change the polarization of surface waves [Fig. 2(c, d)]; see also Fig. S4. In what follows, we shall operate at 2.5 THz, the frequency with TE-TM mode accidental degeneracy.

We now proceed to demonstrate the active control of near-field directionality through tailoring the polarization of surface waves. Here the dipolar source is placed in the middle of two parallel and well-separated graphene-metasurface waveguides [Figs. 3-4]. Therefore, the surface waves in these two waveguides have negligible mutual coupling.

When the source is the bi-state dipole in Fig. 1(a), TM surface waves will be excited and propagate primarily to two opposite $\pm x$ directions in these parallel waveguides if $\mu_c = 0.3$ eV [Fig. 3(a)]. This phenomenon is a hallmark of circular electric dipoles in the near-field coupling [5,25]. In contrast, if $\mu_c = 0$ eV, the excited TE surface waves would propagate along the same $-y$ direction in these two waveguides [Fig. 3(b)]. This phenomenon shows the feature of Huygens dipoles in the near-field coupling [25]. Similarly, Fig. 3(c, d) shows the field distribution of the excited surfaces waves in these two graphene-metasurface waveguides when the source is the bi-state dipole in Fig. 1(b). If $\mu_c = 0.3$ eV, the excited TM surface waves now mainly propagate along the same $+y$ direction in these two waveguides [Fig. 3(c)], indicating that the bi-state dipole in the near-field coupling is a Huygens dipole. In contrast, if $\mu_c = 0$ eV, TE surface waves are excited and propagate along two opposite $\pm x$ directions in these two waveguides [Fig. 3(d)]. Therefore, the bi-state dipole in the near-field coupling becomes to a circular magnetic dipole. Figure 3 indicates that the near-field directionality of the bi-state dipoles is influenced by the polarization of surface waves they couple to. Remarkably, these bi-state dipoles can be actively switched from a circular electric/magnetic dipole to a Huygens dipole by simply tailoring the polarization of surface waves.

We now discuss the possibility to actively control the near-field directionality of Janus dipoles [Fig. 4]. According to the definition [25], the orthogonal electric and magnetic dipole components for Janus dipoles, such as $m_y = icp_0$ and $p_x = p_0$ in Fig. 4, have magnitudes satisfying the Kerker's condition and a 90º phase difference [25]. A critical feature is that only the guided surface waves in one waveguide can be efficiently excited if we put the Janus dipole in the middle of two parallel waveguides (e.g., the graphene-metasurface waveguides) [Fig. 4]. In other words, the Janus dipole has two opposite faces in the near-field coupling, namely a coupling face and a non-coupling face. If $\mu_c = 0.3$ eV, only the TM surface waves in the top waveguide are efficiently excited [Fig. 4(a)]. This phenomenon indicates that the coupling face of Janus dipole is upward for TM surface waves [Fig. 4(a)]. On the contrary, if $\mu_c = 0$ eV, only the TE surface waves in the bottom waveguide are excited, indicating the coupling face of Janus dipoles is downward for TE surface waves [Fig. 4(b)]. Therefore, Fig. 4 shows that the coupling face of Janus dipoles can be actively flipped through tailoring the polarization of surface waves.

Last but not least, we show that the realistic material loss would not deteriorate the near-field directionality [Figs. S5-S6], while it may largely reduce the propagation length of excited surface waves. On the other hand, the dipoles in Figs. 3-4 can be further optimized to achieve the perfect near-field directionality [Figs. S7-S8]. In addition, the active near-field directionality can also occur at the frequency having $k_\parallel^{TM} \neq k_\parallel^{TE}$ [Fig. S9], except for the frequency having $k_\parallel^{TM} = k_\parallel^{TE}$.

## 3. Conclusion

In conclusion, we have revealed that the near-field directionality is determined both by the property of dipolar sources and the polarization of surface waves. Due to the importance of the polarization of surface waves in the near-field coupling, we have proposed a feasible scheme via toggling the polarization of surface waves to achieve the active near-field directionality. Particularly, by applying this scheme in the near-field coupling, a circular electric/magnetic dipole can be actively transformed into a Huygens dipole, and moreover, the near-field coupling and non-coupling faces of Janus dipoles can be actively flipped. The active tunability for these elemental dipolar sources (i.e., circular, Huygens, and Janus dipoles) in the near-field coupling should be desirable for modern photonic applications, especially in the design of on-chip active photonic devices. Due to the flexibility in manipulating the polarization of surface waves, the graphene-metasurface waveguides provide a versatile platform for the active near-field directionality and the design of active optical functionalities.

**Supporting Information**
Supporting Information is available from the Wiley Online Library or from the author.

**Acknowledgements**
The work was sponsored by NSF/EFRI-1741660, the National Natural Science Foundation of China (NNSFC)

under Grants No. 61625502, No.11961141010, and No. 61975176, the Top-Notch Young Talents Program of China, the Fundamental Research Funds for the Central Universities, and the Singapore Ministry of Education (Grant No. MOE2018-T2-1-022 (S) and MOE2016-T3-1-006).

**Conflict of Interest**
The authors declare no conflict of interest.

**Keywords:**
spin-orbit interaction of light, plasmonics, graphene, metasurface

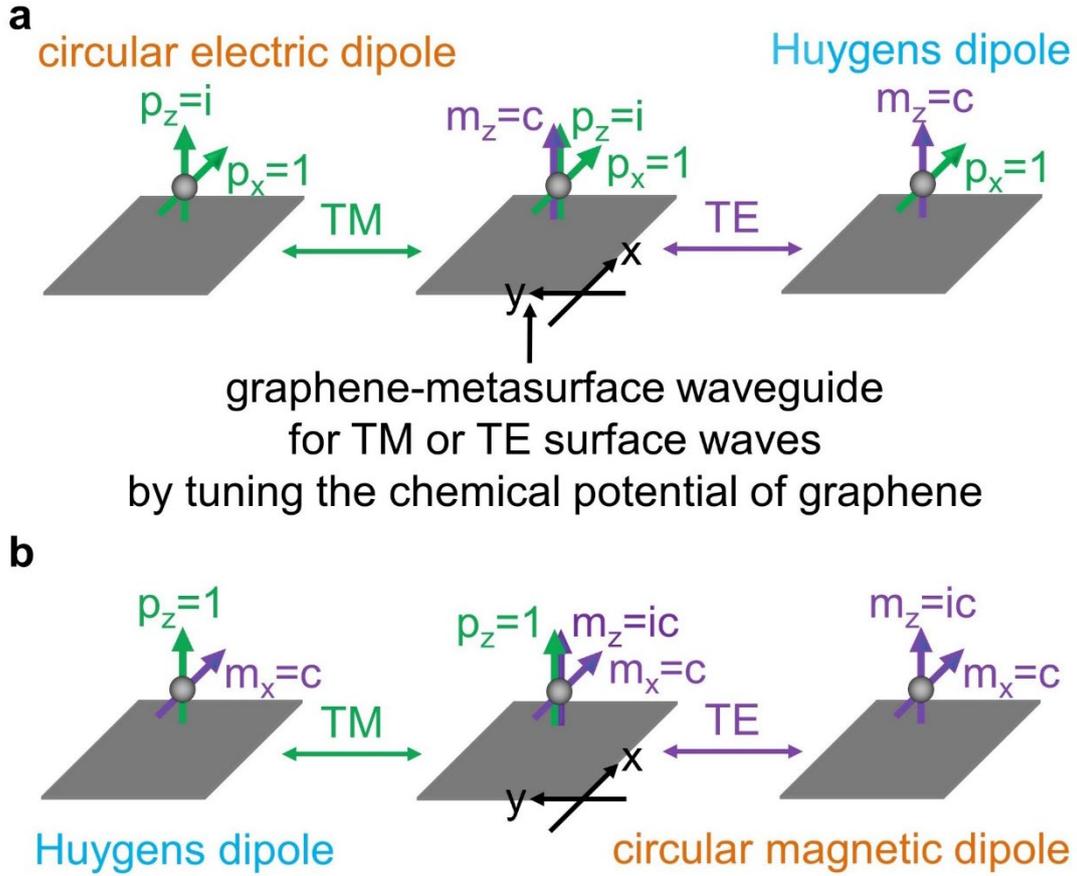

FIG. 1. Exploring the essential role of the polarization of surface waves in the near-field coupling. A dipolar source is placed above a planar waveguide, which supports either TM (*p*-polarized) or TE (*s*-polarized) surface waves. The polarization of surface waves can be actively tailored, for example, via changing the chemical potential of graphene in the graphene-metasurface waveguide. (a, b) For conceptual illustration, we construct two dipolar sources, which have two distinct states in the near-field coupling for different polarizations of surface waves.

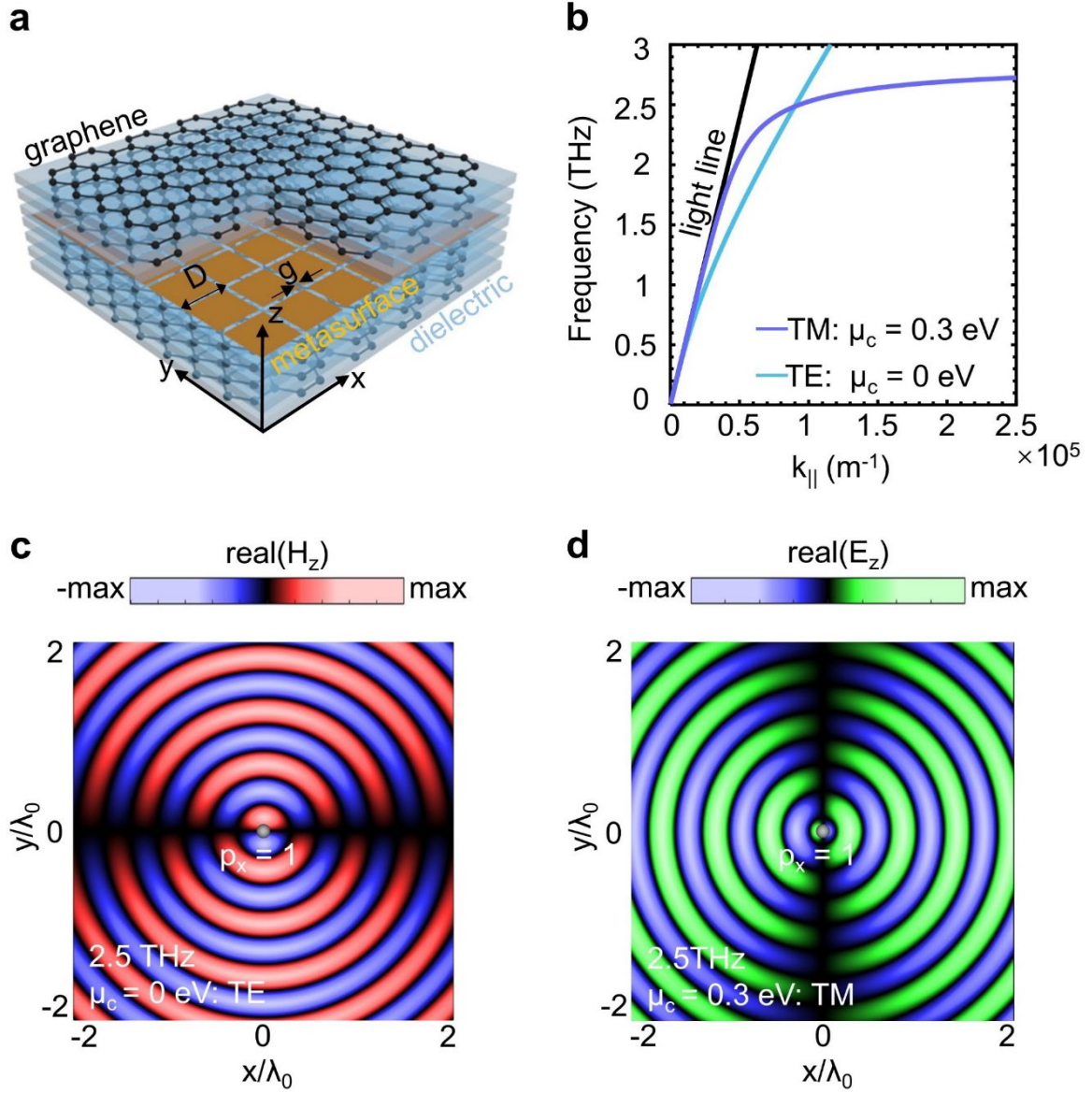

FIG. 2. Tailoring the polarization of surface waves in a graphene-metasurface waveguide. (a) Structural schematic of the waveguide, which consists of one layer of metal-based metasurface and seven layers of graphene. (b) Dispersion of surface waves. The graphene-metasurface waveguide supports TM surface waves if $\mu_c = 0.3$ eV but TE surface waves if $\mu_c = 0$ eV. (c, d) Distribution of excited surface waves from an electric dipole with $\bar{p} = \hat{x} p_x$, whose vertical distance from the waveguide is $0.2\lambda_0$. The working wavelength in free space is $\lambda_0 = 120$ μm in (c, d). The results in (c, d) explicitly show that the near-field coupling is sensitive to the polarization of surface waves.

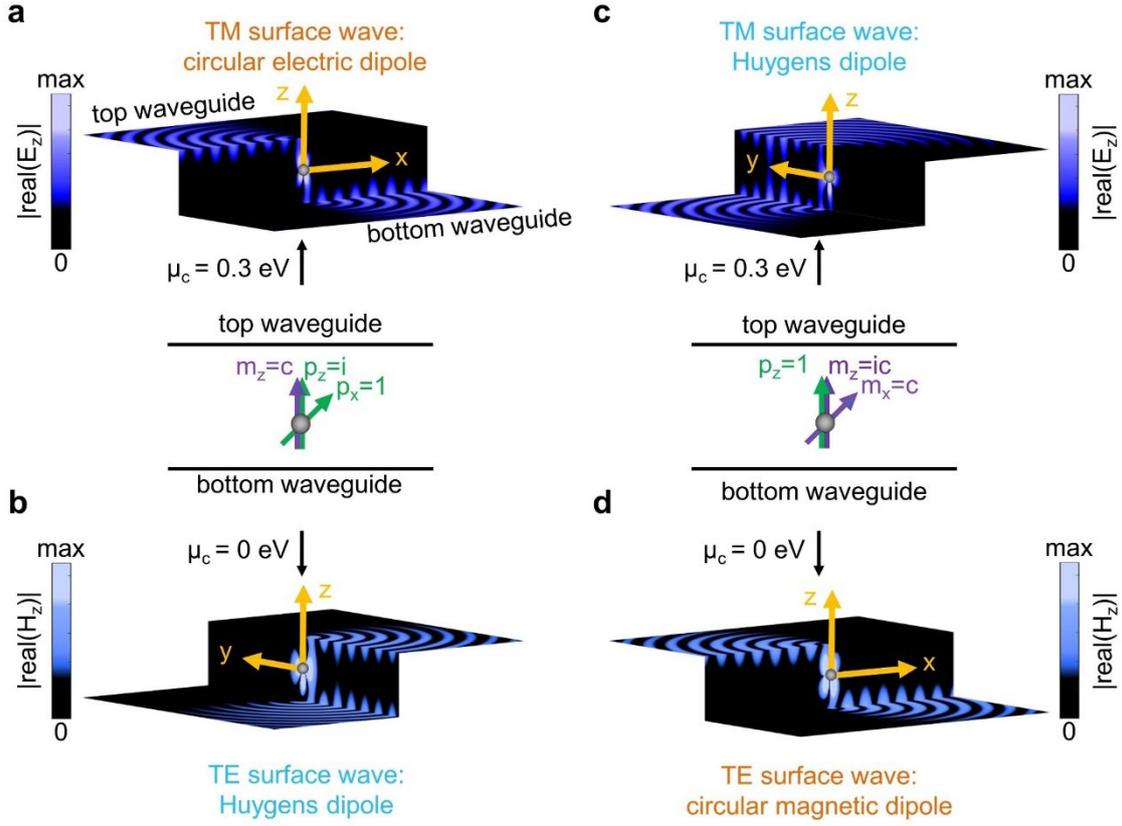

FIG. 3. Manipulating the near-field directionality of dipolar sources (e.g., bi-state dipoles in Fig. 1) by tailoring the polarization of surface waves. The dipolar source is placed in the middle of two parallel and well-separated graphene-metasurface waveguides. (a, b) In the near-field coupling, the dipolar source in Fig. 1(a) is equivalent to a Huygens dipole for TE surface waves if $\mu_c = 0$ eV but becomes to a circular electric dipole for TM wave surface if $\mu_c = 0.3$ eV. (c, d) The dipolar source in Fig. 1(b) behaves like a Huygens dipole for TM surface waves if $\mu_c = 0.3$ eV but transforms into a circular magnetic dipole for TE wave surface if $\mu_c = 0$ eV. The vertical distance between these two waveguides is $0.4\lambda_0$. The working wavelength in free space is $\lambda_0 = 120$ μm. In each plot, the field distributions in the two parallel planes are plotted at the top surfaces of the bottom waveguide and the bottom surface of the top waveguide, respectively.

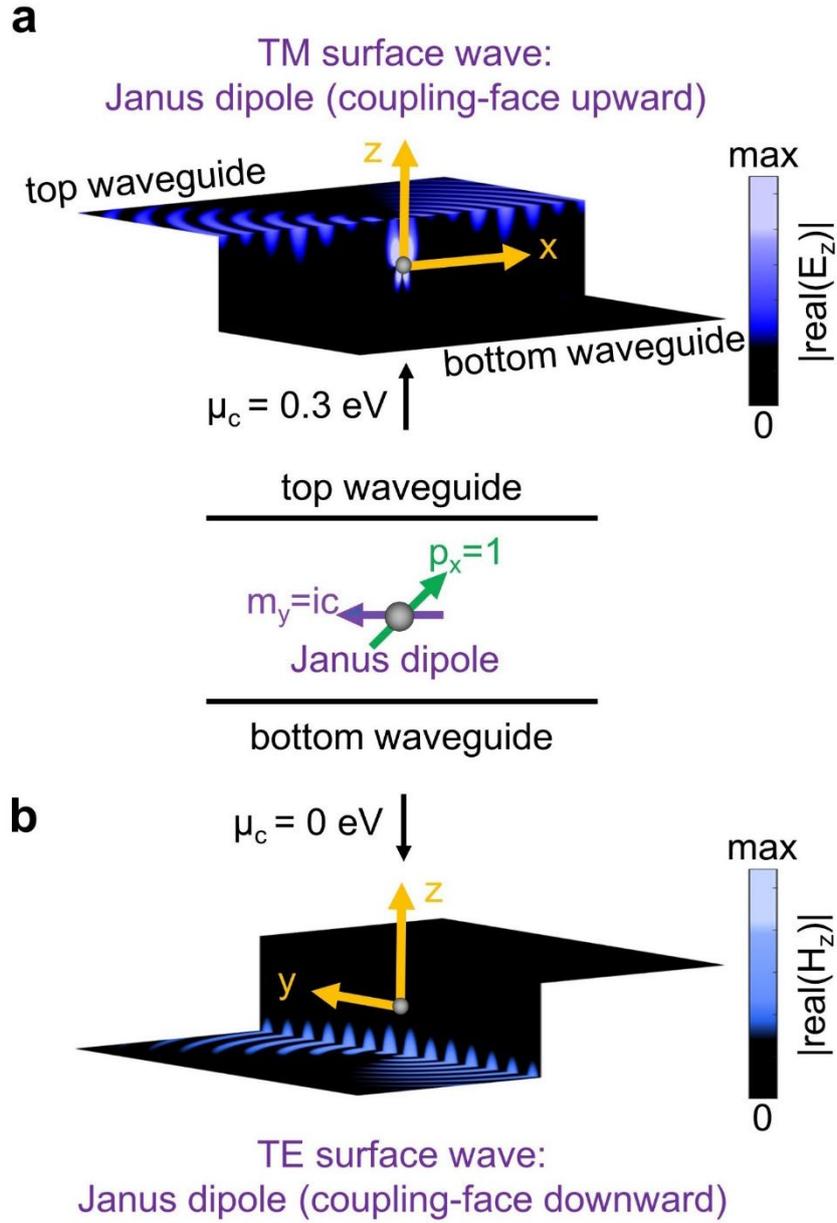

FIG. 4. Actively flipping the coupling and non-coupling faces of Janus dipoles in the near-field coupling. The coupling face of Janus dipoles is facing upward for TM surface waves if $\mu_c = 0.3$ eV but downward for TE surface waves if $\mu_c = 0$ eV. The basic structural setup is the same as Fig. 3, except for the excitation source.